\documentstyle[12pt,epsfig]{article}
\begin{document} 
\begin{center}{\Large \bf One-parameter Darboux transformations in 
Thermodynamics}\end{center}

\bigskip

\centerline{Haret C. Rosu\footnote{E-mail: hcr@ipicyt.edu.mx 
\\Contribution at the AIP 1st International Conference on 
Quantum Limits to the Second Law, San Diego, CA, (main organizer: D.P. Sheehan), end of July 2002,
tentative publication schedule: Dec. 2002.
}}
\centerline{Applied Mathematics Dept., IPICyT, 
Apdo Postal 3-74 Tangamanga,} \centerline{ San Luis Potos\'{\i},
SLP, Mexico}

  
\pagestyle{empty}

\begin{abstract}
{\scriptsize
\noindent
The quantum oscillator thermodynamic actions are the conjugate intensive parameters for the frequency in any 
frequency changing process. These oscillator actions fulfill simple Riccati equations.
Interesting Darboux transformations of the fundamental
Planck and pure vacuum actions are discussed here in some detail. 
It is shown that the one-parameter ``Darboux-Transformed-Thermodynamics" refers
to superpositions of boson and fermion excitations of positive and negative 
absolute temperature, respectively.
A Darboux generalization of the fluctuation-dissipation theorem is also briefly sketched.}
\end{abstract}

\vspace{5mm}



Using the Planck energy distribution of a quantum oscillator one can define an action function
\begin{equation} \label{2}
f_{P}(x)=\frac{U_P}{\omega}=\frac{\hbar}{2}+\frac{\hbar}
{\exp (\hbar x)-1}=\frac{\hbar}{2}{\rm coth}\left(\frac{\hbar x}{2}\right)~,
\end{equation}
where $x=\beta \omega$.
This action plays the role of a generalized force in any process of
frequency change performed by the oscillator.

The point now is that the function
$f_{P}(x)$ fulfills as particular solution the following Riccati equation
\begin{equation} \label{4}
\frac{df}{dx}+f^2=\left(\frac{\hbar}{2}\right)^2~.
\end{equation}
As a matter of fact, the constant vacuum action $f_V=\frac{\hbar}{2}$ is also
a solution of the same Riccati equation, whereas the
pure thermal action
$f_T=\frac{\hbar}{\exp (\hbar x)-1}$ is a particular solution of 
the following equation
\begin{equation} \label{5}
\frac{df}{dx}+\hbar f +f^2=0~,
\end{equation}
which is again a particular type of Riccati equation (a Bernoulli equation). Since the Riccati equation is a main ingredient 
in supersymmetric quantum mechanics \cite{review} we would like to present here some thermodynamic consequences of using 
methods belonging to the area which focus on  one-parameter Darboux transformations of thermodynamic actions.




In supersymmetric quantum mechanics \cite{W}, which is a simple 
application of Darboux transformations \cite{review}, one usually starts with a 
known particular solution of a Riccati equation without linear term
\begin{equation} \label{6}
\frac{df_{p}}{dx}+f^2_{p}=V_1(x)
\end{equation}
that we call the bosonic Riccati equation. We are not interested here in 
a so-called factorization 
constant that can be placed at the right hand side. The function $V_1$ is an 
exactly solvable potential for the (``imaginary-time") Schr\"odinger equation at zero energy
\begin{equation}\label{7}
\left(\frac{d}{dx}+f_p\right)\left(\frac{d}{dx}-f_p\right)w_b=
\frac{d^2w_b}{dx^2}-V_1(x)w_b=0~.
\end{equation}
The particular solution $w_b$ is usually called a bosonic zero mode. 
It is connected to the Riccati
solution through $f_p=\frac{1}{w_b}dw_b/dx$.
Next, changing the 
sign of the first derivative in Eq.~(\ref{6}) one calculates the outcome 
$V_2(x)$ using the same Riccati solution
\begin{equation} \label{6a}
-\frac{df_{p}}{dx}+f^2_{p}=V_2(x)~.
\end{equation}
We call the latter equation the fermionic Riccati equation.
The function $V_2(x)$ is known as the supersymmetric 
partner of the initial potential $V_1(x)$. The corresponding zero mode fulfills
\begin{equation}\label{7a}
\left(\frac{d}{dx}-f_p\right)\left(\frac{d}{dx}+f_p\right)w_f=
\frac{d^2w_f}{dx^2}-V_2(x)w_f=0~.
\end{equation}

On the other hand, already in 1984, Mielnik \cite{M} studied
the ambiguity of the factorization of the Schr\"odinger equation for the 
oscillator that led him to the general (one-parameter-dependent) 
solution of the companion Riccati equation for that case. In other words,
one looks for the general solution \cite{review} 
\begin{equation} \label{8}
  - \frac{d f_{g}}{dx}+ f_{g}^2 =-\frac{d f_p}{dx}+f^2_p=V_2(x)~.
\end{equation}
The latter equation can be solved for $f_{g}$ by employing the Bernoulli ansatz
$ f_{g}(x)= f_p(x) - \frac{1}{v(x)}$, where $v(x)$ is 
an unknown function \cite{Z}. One obtains for the function $ v(x)$ the
following Bernoulli equation
\begin{equation} \label{9}
 \frac{dv(x)}{dx} + 2v(x)\, f_p(x) =1
\end{equation}
that has the solution
\begin{equation} \label{10}
 v(x)= \frac{{\cal I}_{0b}(x)+ \lambda}{w_b^{2}(x)}~,
\end{equation}
where $ {\cal I}_{0b}(x)= \int _{0}^{x} \,
w_b^2(y)\, dy$,
and we consider $\lambda$ as a positive integration constant that is 
employed as a free parameter.

Thus, the general fermionic Riccati solution is a one-parameter function
$ f_{g}(x; \lambda)$ of the following form
\begin{equation} \label{13}
 f_{g}(x;\lambda)=  f_p(x) -
 \frac{d}{dx}
\Big[ \ln({\cal I}_{0b}(x) + \lambda) \Big]=  \frac{d}{dx}
\Big[ \ln \left(\frac{w_b(x)}{[{\cal I}_{0b}(x) +
\lambda]}\right)\Big]=\frac{d}{dx}\ln w_b(x;\lambda)~
\end{equation}
where $w_b(x;\lambda)=\frac{w_b(x)}{{\cal I}_{0b}(x) +\lambda}$.
The range of the $\lambda$ parameter is conditioned by ${\cal I}_{0}(x) +
\lambda \neq 0$ in order to avoid singularities. This is a well-known 
restriction \cite{M}.
According to the supersymmetric construction one can use
this general fermionic Riccati solution to calculate a one-parameter family
of bosonic potentials as follows
\begin{equation} \label{gen}
\frac{df_{g}}{dx}+f_g^2=V_{1,g}~,
\end{equation}
where 
\begin{equation} \label{pot}
V_{1,g}=V_1-2\frac{d^2}{dx^2}\ln\left({\cal I}_{0b}(x) +\lambda\right)
\end{equation}
enters the linear equation
\begin{equation}\label{Sgen}
\left(\frac{d}{dx}+f_g\right)\left(\frac{d}{dx}-f_g\right)w_b(x;\lambda)=
\frac{d^2w_b(x;\lambda)}{dx^2}-V_{1,g}w_b(x;\lambda)=0~.
\end{equation}

In the limit $\lambda 
\rightarrow \infty$, Eq.~(\ref{gen}) goes into Eq.~({\ref{6}) 
because $f_g\rightarrow f_p$ and $V_{1,g}\rightarrow V_{1}$. 

One can think of Eq.~(\ref{gen}) as a 
generalization of the thermodynamic Riccati equation (\ref{4}). Of 
interest are the one-parameter oscillator actions $f_g$ rather than the 
`potentials' $V_{1,g}$. Since in supersymmetric quantum mechanics $V_{1,g}$
are the general Darboux-transformed potentials, we shall call the functions $f_g$ as
Darboux-transformed actions.   


Going now back to Eq.~(\ref{4}) and using $f=w'/w$, 
where the 
prime denotes the derivative with respect to $x$, one gets the linear equation
\begin{equation}
w^{''}-\left(\frac{\hbar}{2}\right )^2w=0~,
\end{equation}
having the particular zero-mode solution $w_a=W_a\sinh(\frac{\hbar}{2}x)$.
Thus, the general Riccati solution is a one-parameter function
$ f_{gP}(x; \lambda)$ of the following form
\begin{equation}
 f_{gP}(x;\lambda)=  f_P(x) - 
 \frac{d}{dx}
\Bigg[ \ln({\cal I}_{0a}(x) + \lambda) \Bigg]
=  \frac{d}{dx}
\Bigg[ \ln \left(\frac{w_{a}(x)}{[{\cal I}_{0a}(x) +
\lambda]}\right)\Bigg].
\end{equation}


For the vacuum case the  
particular Riccati solution is the vacuum action $f_p=f_V=\frac{\hbar}{2}$. The 
corresponding zero mode is $w_V\propto e^{\hbar x/2}$. 
The usual entropy 
of the vacuum fluctuations is zero whereas the 
modified entropy based on $f_{gV}$ has a kink-like behavior between the $\frac{\hbar}{2}$
(bosonic) solution and the $-\frac{\hbar}{2}$ (fermionic) 
solution. 


One can also use the general zero-mode $w_g=Ae^{\hbar x/2}+Be^{-\hbar x/2}$ 
as solution in Eq.~(\ref{7}). 
The Planck action corresponds to $A=-B=\frac{1}{2}$ (antisymmetric zero-mode), 
while the vacuum case to $A=$ arbitrary and $B=0$. 
A very interesting case corresponds to the symmetric zero-mode, $A=B=\frac{1}{2}$,
$w_s=W_s\cosh (\frac{\hbar}{2} x)$, since if we trace back to the action we get 
\begin{equation}
f_s=-\frac{\hbar}{2} +\frac{\hbar}{\exp (-\hbar x)+1}~,
\end{equation}
i.e., a Fermi-Dirac action for negative $x$.
We have shown in a previous paper that
the $\lambda$ parameter is equivalent to the quotient $A/B$ \cite{boya}.
Thus, the general Riccati solution introduces effects of the second 
linear independent solution. The problem then turns into a subtle 
interpretation of the mathematical results. 
We have found at least one possible physical significance.
For the Planck case, the second linear independent zero-mode is exactly
the $\cosh$ function and the corresponding action is the aforementioned Fermi-Dirac action
of negative $x$. 
We attach the minus sign to the temperature parameter
in $x=\omega /T$ rather than to $\omega$ and recall that 
the issue of negative absolute (spin) temperatures first appeared in Physics in 
1951 when Purcell and Pound were able to produce sudden reversals of the 
direction of an external magnetic field applied to a crystal of LiF \cite{PP}.
Since then many other experiments with negative temperatures have been
devised in nuclear spin systems and the corresponding  
`violations' of the second law of thermodynamics 
were a subject of discussion \cite{ram}.
Thus, the one-parameter Darboux transformations of the Planck action
are a way of introducing 
the effect of a Fermi-Dirac action of negative absolute temperature.


 
Another interesting application refers to dissipative $RLC$ systems 
where the oscillator action enters
Nyquist-Johnson spectral power noise distributions of the type 
(fluctuation-dissipation theorem \cite{N})
\begin{equation}
P(\omega ,\beta)=\frac{\omega}{\pi}R(\omega ,\beta)f_P~.
\end{equation}
One can think of the corresponding generalization
\begin{equation}
P(\omega , \beta ;\lambda)=\frac{\omega}{\pi}R(\omega ,\beta)f_g(\beta \omega;
\lambda)
\end{equation}
and hope to study even in simple experiments the significance of this Darboux generalization of 
the fluctuation-dissipation theorem.



In conclusion, Planck's thermodynamic oscillator action  
can be generalized to a one parameter Darboux
family of actions. We also consider the bosonic vacuum case separately in 
the same way.
The Planck action and the pure vacuum case correspond to the asymptotic 
limit of the Darboux parameter $\lambda \rightarrow \infty$. 
In the Planck case, all the other $\lambda$
cases correspond to a system made of bosons at temperature $T$
interacting with an equal system of fermions at temperature $-T$.
In the vacuum case, the $\lambda \neq \infty$ cases describe the interaction
of the bosonic and fermionic vacua. 
Using the procedure proposed by  Arnaud, Chusseau and Philippe \cite{acp} we have calculated 
the efficiencies of the ideal oscillator Carnot
cycles based on the Darboux-modified Planck and vacuum entropies in a previous work \cite{rac}. 
Systems of negative Kelvin temperatures are
hotter than those of positive $T$ \cite{ram} and therefore they always 
represent the hot bath. In real, dissipative cases, the same type
of generalization is suggested for the fluctuation-dissipation theorem. 
Finally, we mention that a multiple-parameter Darboux generalization is also
possible \cite{mult}. The experimental nature of the $\lambda$ parameter is an open issue.

\nopagebreak






\nopagebreak


\bigskip


\newpage



\end{document}